\documentclass[sigplan,screen]{acmart}
\setcopyright{none}
\renewcommand\footnotetextcopyrightpermission[1]{}
\usepackage{multirow}

\usepackage{graphicx}
\usepackage[caption=false]{subfig} 
\usepackage{hyperref}
\usepackage{caption}

\let\cref\autoref
\let\Cref\Autoref
\makeatletter
\renewcommand{\p@subfigure}{\figurename~\thefigure}
\makeatother

\usepackage{enumitem}
\usepackage{dblfloatfix}

\AtBeginDocument{%
  }

\acmConference[]{}{}

\begin{document}

\title{Real-Time Per-Garment Virtual Try-On with Temporal Consistency for Loose-Fitting Garments}

\author{Zaiqiang Wu}
\affiliation{%
 \institution{The University of Tokyo}
 \country{Japan}
}
\author{I-Chao Shen}
\affiliation{%
 \institution{The University of Tokyo}
 \country{Japan}
}
\author{Takeo Igarashi}
\affiliation{%
 \institution{The University of Tokyo}
 \country{Japan}
}

\renewcommand{\shortauthors}{Wu et al.}


\begin{abstract}
Per-garment virtual try-on methods collect garment-specific datasets and train networks tailored to each garment to achieve superior results. However, these approaches often struggle with loose-fitting garments due to two key limitations: (1) They rely on human body semantic maps to align garments with the body, but these maps become unreliable when body contours are obscured by loose-fitting garments, resulting in degraded outcomes; (2) They train garment synthesis networks on a per-frame basis without utilizing temporal information, leading to noticeable jittering artifacts. To address the first limitation, we propose a two-stage approach for robust semantic map estimation. First, we extract a garment-invariant representation from the raw input image. This representation is then passed through an auxiliary network to estimate the semantic map. This enhances the robustness of semantic map estimation under loose-fitting garments during garment-specific dataset generation. To address the second limitation, we introduce a recurrent garment synthesis framework that incorporates temporal dependencies to improve frame-to-frame coherence while maintaining real-time performance. We conducted qualitative and quantitative evaluations to demonstrate that our method outperforms existing approaches in both image quality and temporal coherence. Ablation studies further validate the effectiveness of the garment-invariant representation and the recurrent synthesis framework.

\end{abstract}




\begin{teaserfigure}
\centering
    \subfloat[Real-time virtual try-on]{%
        \includegraphics[height=0.33\linewidth]{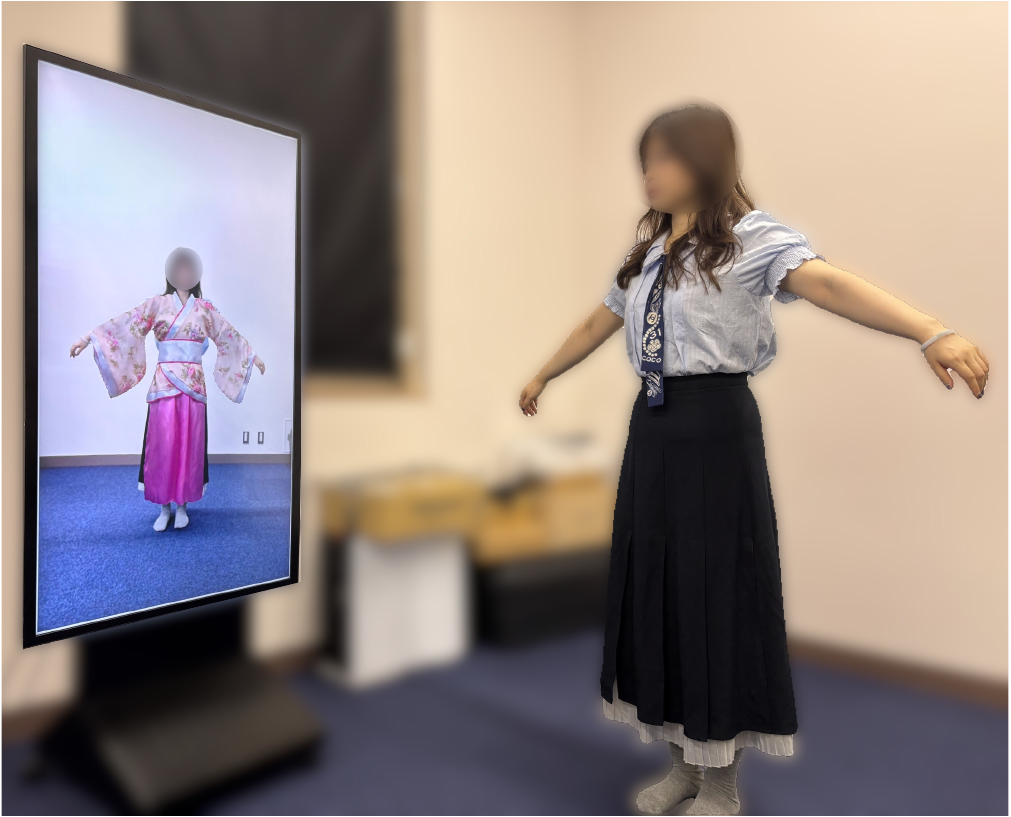}
    }
    \hfill
    \subfloat[Loose-fitting garment try-on results]{%
        \includegraphics[height=0.33\linewidth]{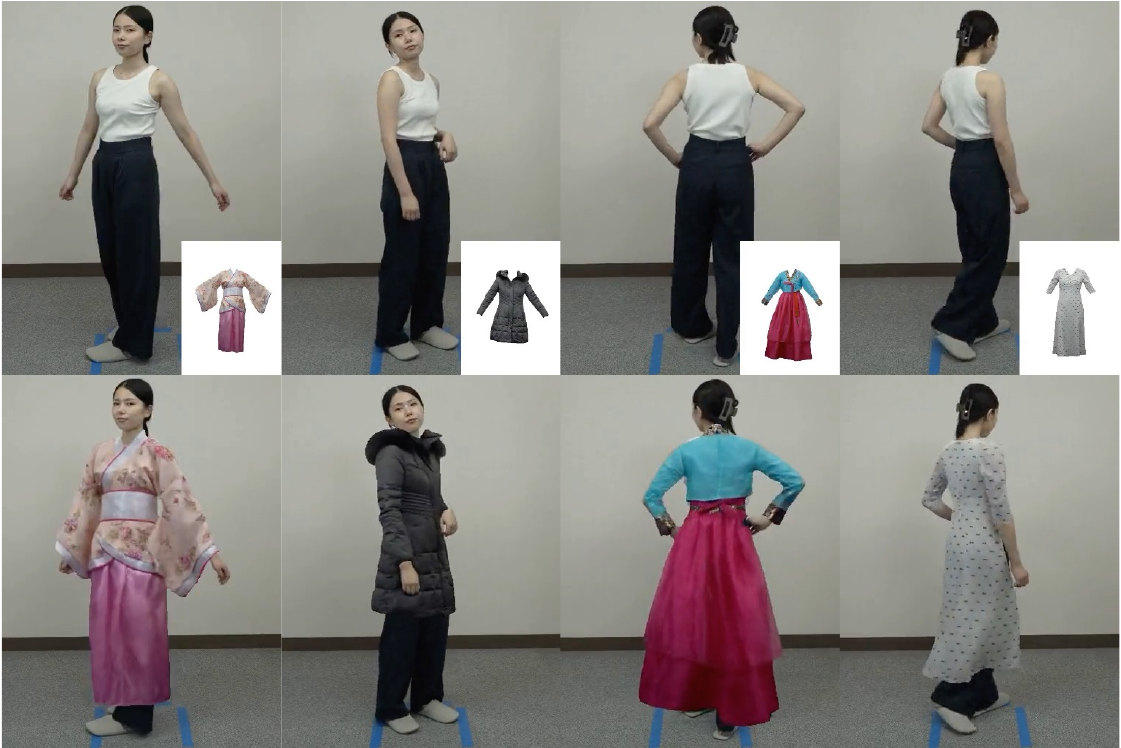}
    }
    \caption{(a) Our virtual try-on system features an efficient inference pipeline that enables real-time performance. (b) Our method produces highly realistic $360^{\circ}$ virtual try-on results for loose-fitting garments.}
\label{fig:loose_teaser}
\end{teaserfigure}


\maketitle

\begin{figure}[htb]
\centering 
\includegraphics[width=\linewidth]{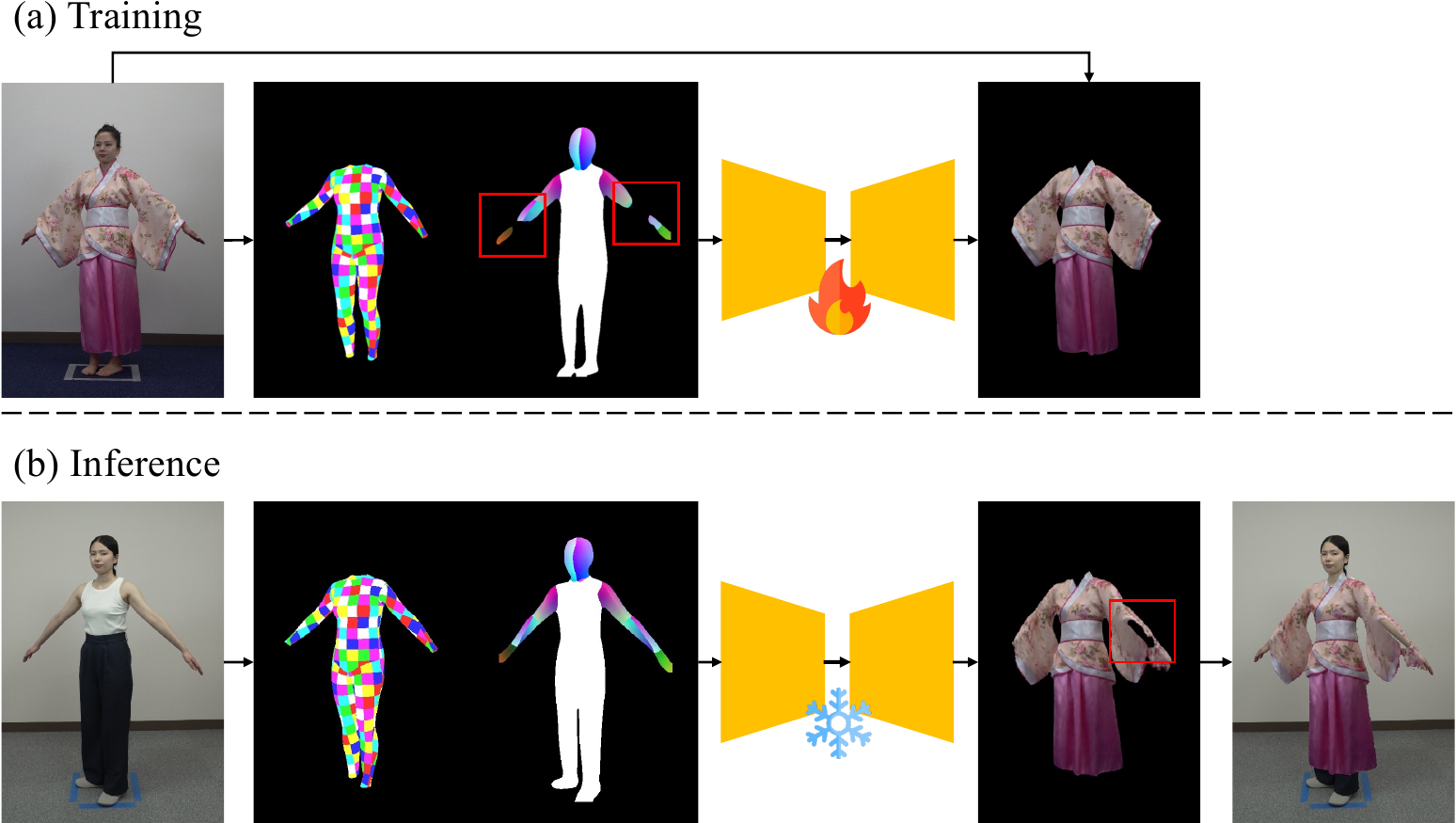}
  
  \caption{Illustration of the training and inference pipelines for applying the method proposed in~\cite{wu2025low} to virtually try on a loose-fitting garment. (a) During training, the garment synthesis network is trained using degraded DensePose~\cite{Guler2018DensePose} outputs. (b) During inference, a distribution mismatch between training and input data leads to suboptimal garment synthesis results.}
  \label{fig:loose_discrepancy}
\end{figure}

\section{Introduction}
\label{sec:loose_intro}
Virtual try-on technology, which enables customers to try on garments without physical access, has recently garnered significant attention from researchers in today's e-commerce landscape. Existing image-based virtual try-on methods~\cite{jetchev2017conditional,dong2019fw,choi2021viton,lee2022high,fang2024vivid,karras2024fashionvdmvideodiffusionmodel} require only a 2D in-shop image of the target garment to generate virtual try-on results, making them highly accessible. However, these approaches are predominantly tailored for retailers aiming to produce promotional images featuring fashion models, rather than for end-users seeking personalized virtual try-on experiences. This focus is clearly reflected in the biased datasets used for training, which are heavily skewed toward tall and slim fashion models. As a result, these methods fail to generalize well to the diverse body types of average consumers.

Unlike general image-based methods, per-garment virtual try-on methods~\cite{chong2021per,wu2024virtual,wu2025low} focus on customer-oriented virtual try-on by collecting garment-specific datasets and training garment-specific neural networks. Early per-garment methods~\cite{chong2021per,wu2024virtual} require an expensive robotic mannequin for dataset collection, making their results non-reproducible. A follow-up work \cite{wu2025low} lowers the barrier to dataset collection significantly by using real human bodies for dataset collection. This approach relies on DensePose~\cite{Guler2018DensePose} for estimating human body semantic maps, which are crucial for aligning the human body and the synthesized garment in 2D image space. However, it struggles with \emph{loose-fitting} garments, as DensePose results degrade significantly for these types of clothing, causing a distribution discrepancy between the training and inference stages, as illustrated in~\cref{fig:loose_discrepancy}. Such discrepancies lead to degraded results during the inference stage, where users typically wear tight-fitting garments to facilitate accurate human body tracking.

Moreover, previous per-garment approaches rely solely on the current input frame for garment synthesis, neglecting the \emph{temporal continuity} provided by preceding frames. As a result, their results often exhibit jittering artifacts—especially for loose-fitting garments, which naturally exhibit appearance variations even under identical poses.

To address these limitations, we introduce a garment-invariant representation that can be directly extracted from the raw input image. This representation is then transformed into a human body semantic map using an auxiliary network, enabling robust estimation of the human body semantic map under loose-fitting garments during garment-specific dataset generation. The garment-invariant representation leverages joint heatmaps and 3D human pose, both of which are inherently robust to garment variations. Furthermore, we propose a recurrent garment synthesis framework that ensures temporal consistency by conditioning the synthesis process on both the current input frame and the preceding temporal states.

Our recurrent garment synthesis framework is capable of processing input sequences of arbitrary length while maintaining high efficiency, achieving approximately 10 frames per second on a standard PC. This ensures a smooth and responsive virtual try-on experience for users, as demonstrated in~\cref{fig:loose_teaser}. We performed both qualitative and quantitative comparisons against general image-based methods and the existing per-garment method to demonstrate the superiority of our approach in terms of visual quality and temporal consistency. In addition, we validated the effectiveness of the garment-invariant representation by replacing it with alternative representations and assessed the necessity of the recurrent synthesis framework through ablation studies.

Our contributions can be summarized as follows: 
\begin{itemize} 
	\item We propose a garment-invariant representation for robust human body semantic map estimation under loose-fitting garments, enabling per-garment virtual try-on for such garments. 
	\item We develop a recurrent garment synthesis framework that supports real-time virtual try-on with temporal consistency, accommodating input sequences of arbitrary length.
	\item We conduct comprehensive qualitative and quantitative comparisons with existing methods, along with ablation studies, to validate the effectiveness and necessity of each component in our proposed framework.        
\end{itemize}
\section{Related Work}
\label{sec:related}
\subsection{3D model-based Virtual Try-On}
3D model-based virtual try-on methods~\cite{guan2012drape,sekine2014virtual,santesteban2019learning,patel2020tailornet,pan2022predicting} typically represent garments using polygon meshes. These methods generate try-on results by simulating the interaction between clothing and a 3D human body model, leveraging either physics-based simulations~\cite{cirio2014yarn,kaldor2008simulating,narain2012adaptive,selle2008robust} or learning-based approaches~\cite{casado2022pergamo,grigorev2023hood,halimi2023physgraph,lahner2018deepwrinkles,pan2022predicting,patel2020tailornet,santesteban2019learning,santesteban2022snug,santesteban2021self,xiang2021modeling}. While these techniques can realistically capture garment deformations across diverse poses and viewpoints, they require labor-intensive manual 3D modeling for each garment, which poses a significant barrier to scalability. To address this, some recent approaches~\cite{feng2022capturing,dong2023vica,Li_2024_CVPR,lin2024layga,chen2024gaussianeditor,chen2024gaussianvton,cao2024gs,he2025vton} adopt volumetric representations, such as Neural Radiance Fields (NeRF)~\cite{mildenhall2021nerf} and 3D Gaussian Splatting (3DGS)~\cite{kerbl20233d}, to model garments directly from monocular images. These representations facilitate automatic 3D garment reconstruction but are primarily designed for digital avatars and are not optimized for real-time virtual try-on experiences involving real human users.

In contrast, our method supports real-time virtual try-ons for real-world users and enables automatic pre-processing for each garment with minimal human intervention.

\subsection{Image-based Virtual Try-On}
Image-based virtual try-on methods eliminate the need for 3D modeling of garments, relying instead on a single 2D in-shop image to generate try-on results~\cite{song2023image}. This makes them significantly more accessible and scalable compared to 3D model-based approaches. Several methods~\cite{jetchev2017conditional,choi2021viton,lee2022high,kim2024stableviton,wang2024mv,xu2024ootdiffusion} focus on producing static try-on images, which lack the ability to convey dynamic garment behavior. To address this, other approaches~\cite{dong2019fw,chen2021fashionmirror,jiang2022clothformer,fang2024vivid,xu2024tunnel,karras2024fashionvdmvideodiffusionmodel} incorporate inter-frame consistency to generate temporally coherent video try-on results. However, these methods typically suffer from high memory usage, restricting the length of video they can process. In addition, most image-based methods rely on datasets scraped from online images featuring tall and slim fashion models, which leads to suboptimal performance when applied to average users.

In contrast, our method enables real-time virtual try-on with temporal consistency across arbitrary video lengths and demonstrates strong generalization to unseen body shapes.

\subsection{Per-garment Virtual Try-On}
Unlike image-based virtual try-on approaches that rely on universal networks trained across diverse garment types, per-garment methods focus on training garment-specific networks using dedicated datasets, resulting in superior try-on quality for specific garments. A pioneering effort by~\cite{chong2021per} introduced the use of a robotic mannequin to automate the collection of garment-specific datasets, enabling the training of garment-specific networks. However, this approach requires users to wear a physical measurement garment for accurate body tracking, which limits its practicality. To address this, \cite{wu2024virtual} proposed a vision-based intermediate representation that eliminates the need for the physical measurement garment. More recently, \cite{wu2025low} further reduced the barrier to dataset collection by leveraging real human bodies, thereby removing the dependency on costly robotic mannequins. Despite these advancements, \cite{wu2025low} struggles with loose-fitting garments due to their reliance on human body semantic map estimation, which often fails under such garments, and the jittering artifacts resulting from a lack of inter-frame consistency enforcement.

Our method addresses these limitations by introducing a garment-invariant representation for human body semantic map estimation and a recurrent garment synthesis framework, enabling high-quality dataset generation and temporally consistent virtual try-ons for loose-fitting garments.

\subsection{Domain Adaption}
Domain adaptation aims to ensure the performance of machine learning models trained on a source domain when applied to a target domain with a different data distribution~\cite{farahani2021brief}. The predominant paradigm of domain adaptation is mapping the source and target domains into a shared space~\cite{tzeng2019deep,tzeng2017adversarial,tzeng2015simultaneous,sankaranarayanan2018generate,hoffman2016fcns}. Hoffman et al.~\cite{hoffman2016fcns} present a domain adaptation framework with fully convolutional networks for semantic segmentation. They learn the mapping to the shared space in an unsupervised manner by assuming source and target domains share the same label statistics. Isola et al.~\cite{isola2017image} directly learns the mapping from the source domain to the target domain in a supervised way. However, the correlations between the source and target domains are usually unknown. The works in \cite{royer2020xgan} and \cite{zhu2017unpaired} use cycle consistency to learn the mappings between unpaired source and target images. However, unpaired image-to-image translation struggles with high-resolution images, limiting their applicability in our scenario.

In this paper, we propose a shared representation space for images of humans wearing tight-fitting and loose-fitting garments. Our aim is to bridge the domain gap that causes existing human body semantic map estimation models, trained predominantly on tight-fitting garment images, to underperform when applied to loose-fitting garment images.
\begin{figure*}[htb]
  \centering
  \includegraphics[width=\linewidth]{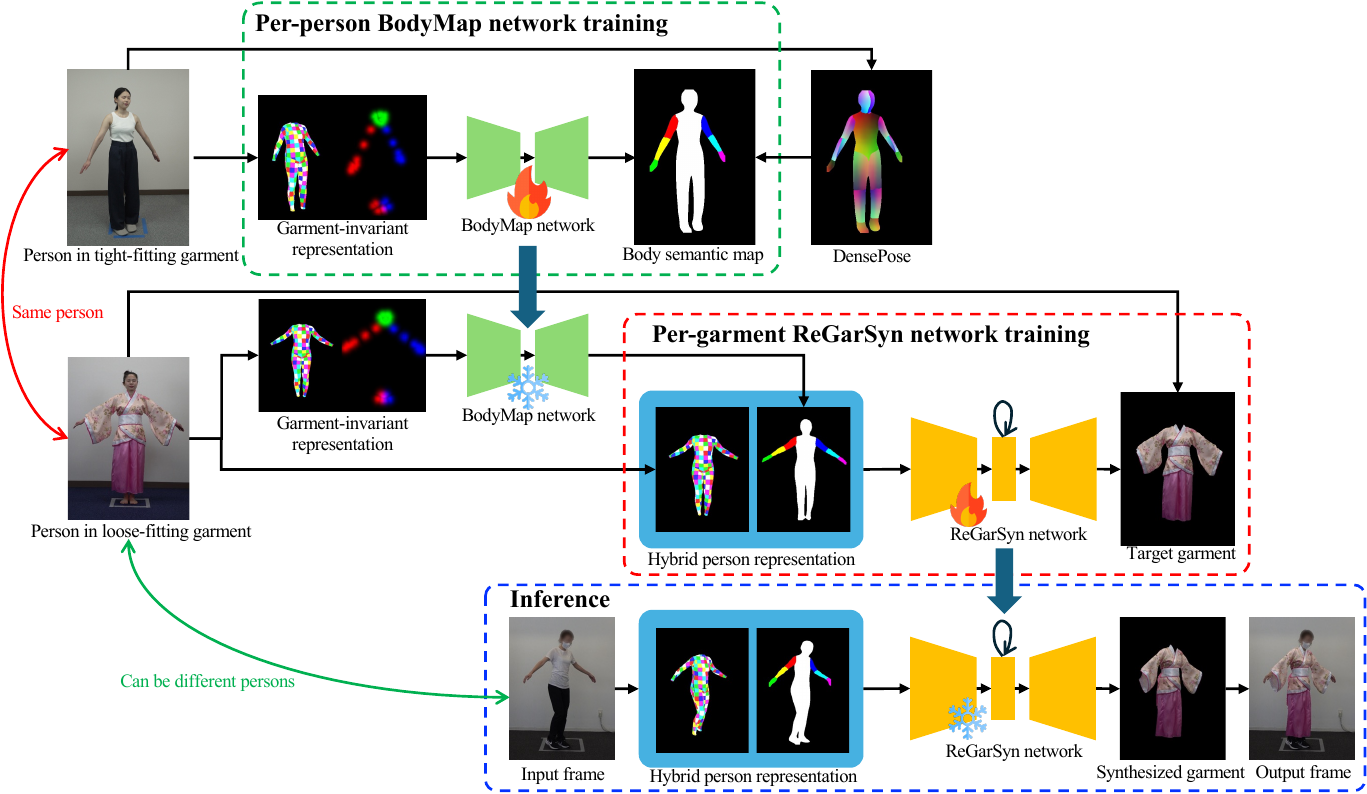}
  \caption{Overview of our method. We first train a per-person BodyMap network using images of a person wearing tight-fitting garments. This network is then used to robustly estimate body semantic maps for the same person in a loose-fitting garment. These estimated maps are subsequently used to train a per-garment ReGarSyn network. Once trained, the ReGarSynt network enables real-time virtual try-on for arbitrary persons.}
  \label{fig:loose_overview}
\end{figure*}

\begin{figure}[htb]
  \centering
  
  \includegraphics[width=.9\linewidth]{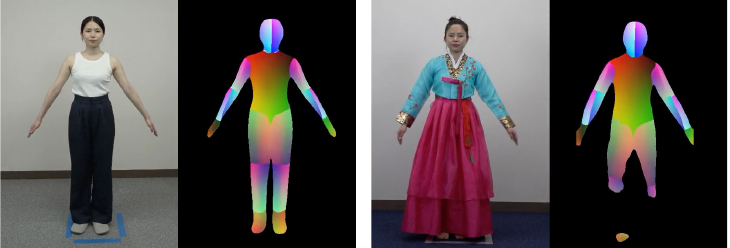}
  
  \caption{DensePose~\cite{Guler2018DensePose} accurately estimates the human body semantic map when subjects wear tight-fitting garments (left). However, its performance significantly degrades with loose-fitting garments due to occlusions (right).}
  \label{fig:loose_densepose}
\end{figure}

\section{Method}

\subsection{Overview}

\cref{fig:loose_overview} provides an overview of our method. To mitigate the distribution gap between the training and inference stages, we aim to obtain robust human body semantic maps under loose-fitting garments. To achieve this, we propose a garment-invariant representation that is robust to garment variations. Using images of a person wearing a tight-fitting garment, we train a per-person BodyMap network to translate this representation into the body semantic map.

Once trained, the BodyMap network can be used to robustly estimate body semantic maps for the same person in the target loose-fitting garment by transforming the extracted garment-invariant representations into body semantic maps. These estimated maps are then used to train a per-garment Recurrent Garment Synthesis (ReGarSyn) network that learns the mapping from the input sequence to the target garment image sequence. 

Unlike prior works~\cite{chong2021per,wu2024virtual,wu2025low}, which synthesize garment images frame by frame, our proposed ReGarSyn network synthesizes garment images in a recurrent manner to enhance temporal consistency. During inference, our recurrent framework facilitates temporally coherent virtual try-on for arbitrary persons over long input sequences while maintaining real-time performance and a constant memory footprint.

While it is technically feasible to use the garment-invariant representation as input to the ReGarSyn network, which eliminates the need for the BodyMap network, we opted against this approach due to concerns regarding inference speed. Generating the garment-invariant representation is computationally intensive, whereas the semantic map can be extracted in real-time. Therefore, it is necessary to train an auxiliary BodyMap network that maps the garment-invariant representation to the semantic map, thereby allowing us to bypass the computational overhead of generating the garment-invariant representation during inference.

\subsection{Design of Garment-Invariant Representation}

\begin{figure}[htb]
    \centering
    \subfloat[Garment's influence on the human joint locations.]{%
        \includegraphics[width=0.9\linewidth]{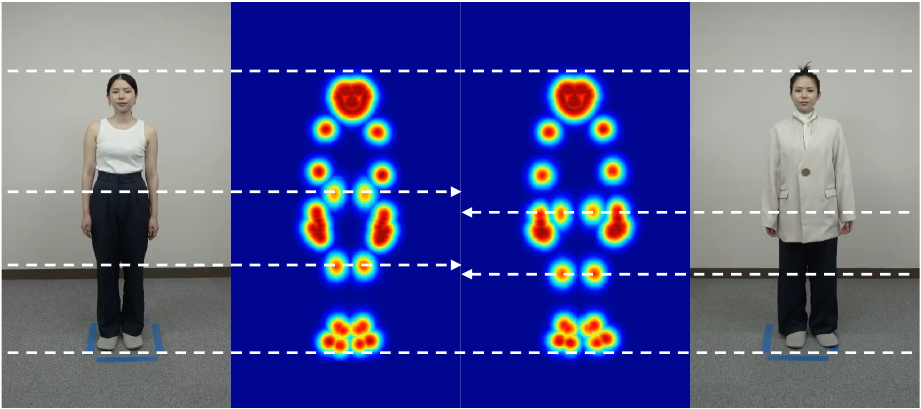}
    }
    \hspace{-0.1cm}
    \subfloat[Garment's influence on human joint confidence scores.]{%
        \includegraphics[width=0.9\linewidth]{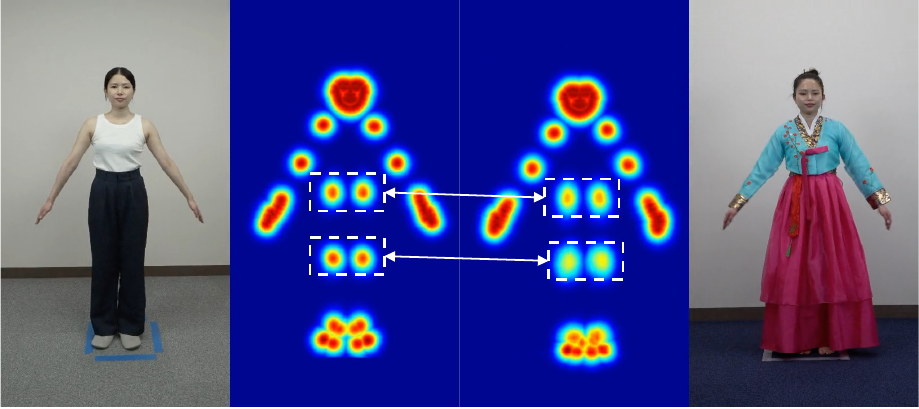}
    }
    \caption{Human joint estimation is not fully garment-invariant, especially for the knee and hip joints, whose positions and confidence scores are sensitive to garment variations.}
    \label{fig:loose_heatmap}
\end{figure}

The prior work \cite{wu2025low} employs DensePose~\cite{Guler2018DensePose} for estimating the human body semantic map directly from the raw input image. However, DensePose varies significantly under loose-fitting garments, as shown in~\cref{fig:loose_densepose}. To ensure reliable human body semantic map estimation under loose-fitting garments, we need a garment-invariant representation that captures human body information and is invariant to the garment to serve as the intermediate representation. A natural candidate for such a representation is the set of human joints. However, not all joints are truly garment-invariant. The presence and type of garments can influence their detected positions and associated confidence scores, as shown in~\cref{fig:loose_heatmap}. The human joint estimation results are generated by Sapiens~\cite{khirodkar2024sapiens}, which is the state-of-the-art human pose estimation method.

Although human joint estimation is not entirely garment-invariant, it can still serve as a robust representation by excluding joints sensitive to garment variations, such as the hips and knees. However, excluding these joints limits the expressiveness of the resulting pose representation. To address this limitation, we augment the remaining joint data with a virtual measurement garment~\cite{wu2025low}, which is a rendered image of the SMPL mesh that encapsulates the full 3D human pose. In contrast to~\cite{wu2025low}, which uses an upper-body template mesh, we utilize a full-body template mesh to facilitate the representation of full-body pose.

\begin{figure}[htb]
  \centering
  \includegraphics[width=.9\linewidth]{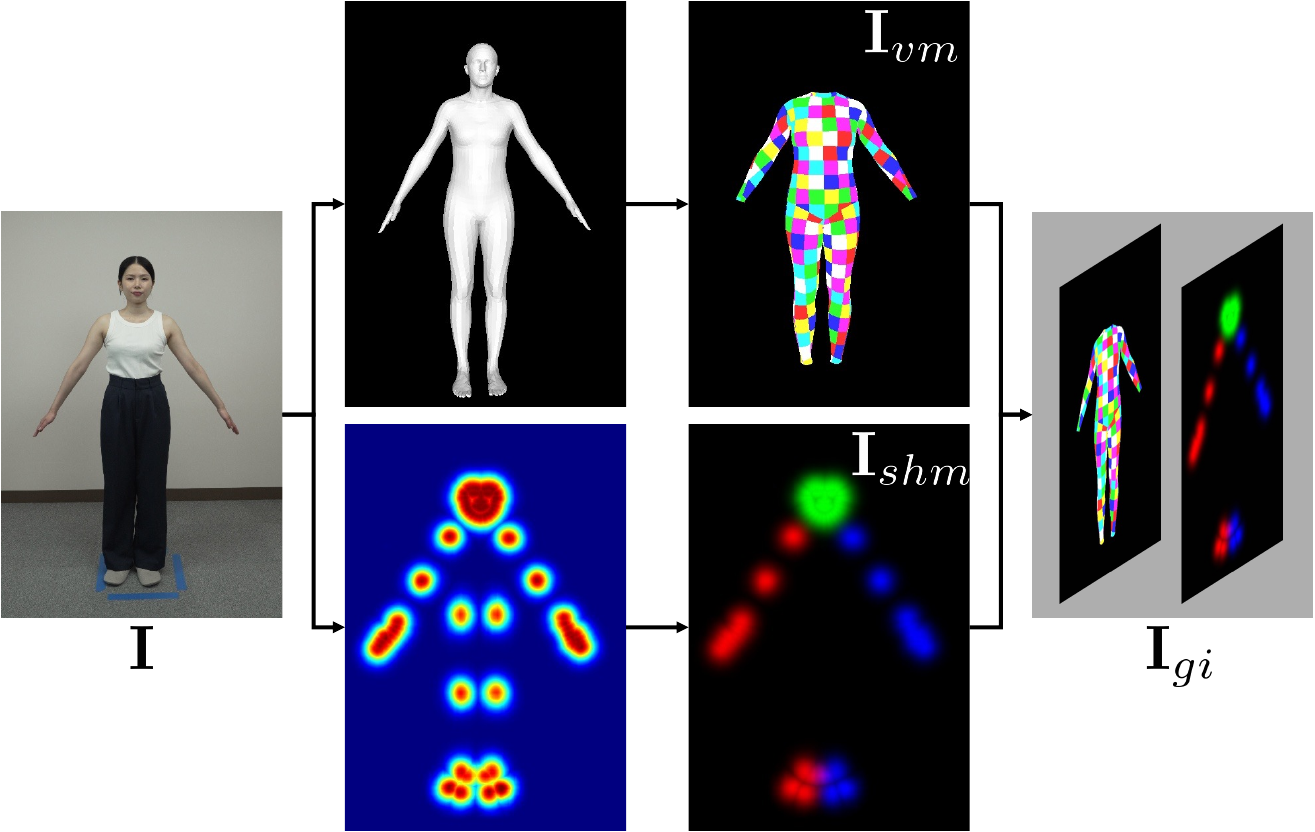}
  \caption{Garment-invariant representation extraction from a person image.}
  \label{fig:loose_garment_invariant}
\end{figure}

As illustrated in~\cref{fig:loose_garment_invariant}, given an input person image $\mathbf{I} \in \mathbb{R}^{3 \times H \times W}$, where $H$ and $W$ denote the height and width respectively, we first estimate the 3D human pose using BEV~\cite{sun2022putting}, which outputs a SMPL mesh~\cite{loper2015smpl}. To ensure garment invariance, we fix the body shape parameters $\mathbf{\beta}\in \mathbb{R}^{10}$, which are sensitive to garment variations, to a zero vector. Subsequently, we remove non-essential body parts (head, hands, and feet) and apply a grid-pattern texture proposed by~\cite{halimi2022pattern} to the remaining mesh, which enhances the expressiveness of human body orientation. This textured mesh is then rendered into an RGB image $\mathbf{I}_{vm} \in \mathbb{R}^{3 \times H \times W}$.

In parallel, we extract 133 joint heatmaps using Sapiens~\cite{khirodkar2024sapiens}. To further reduce sensitivity to garment variation, we discard the heatmaps corresponding to the hip and knee joints. The remaining heatmaps are grouped into three categories: left-side limb joints, right-side limb joints, and head joints. These are compactly represented as a three-channel RGB image, denoted as $\mathbf{I}_{shm} \in \mathbb{R}^{3 \times H \times W}$. As illustrated in \cref{fig:loose_garment_invariant}, the R channel represents right-side limb joints, the G channel encodes head joints, and the B channel corresponds to left-side limb joints.

Combining these components, we construct our proposed garment-invariant representation $\mathbf{I}_{gi} \in \mathbb{R}^{6 \times H \times W}$ as follows:

\begin{align} 
\mathbf{I}_{gi} = \mathbf{I}_{vm} \oplus \mathbf{I}_{shm}
\end{align}

where $\oplus$ denotes channel-wise concatenation.

\subsection{Reliable Human Body Semantic Map Estimation}
\begin{figure}[htb]
    \centering
    \subfloat[]{%
        \includegraphics[width=0.2\linewidth]{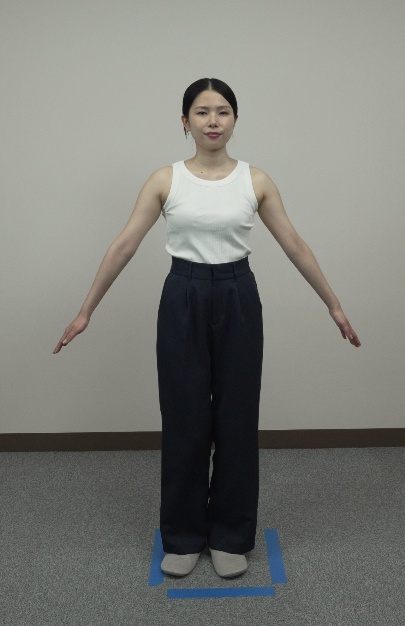}
    }
    \hspace{-0.1cm}
    \subfloat[]{%
        \includegraphics[width=0.2\linewidth]{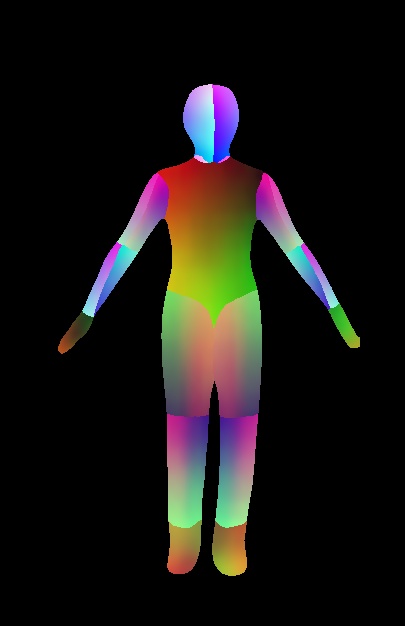}
    }
    \hspace{-0.1cm}
    \subfloat[\cite{wu2025low}]{%
        \includegraphics[width=0.2\linewidth]{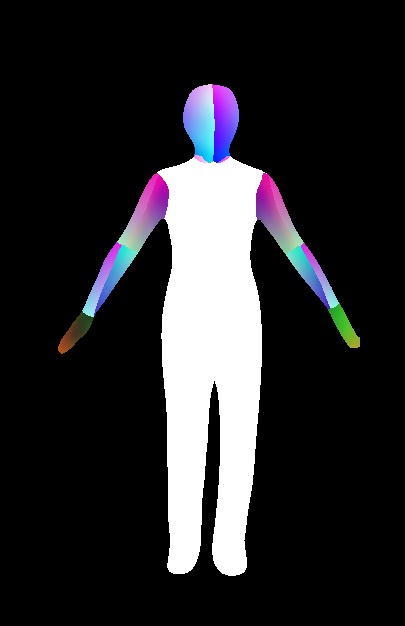}
    }
    \hspace{-0.1cm}
    \subfloat[Ours]{%
        \includegraphics[width=0.2\linewidth]{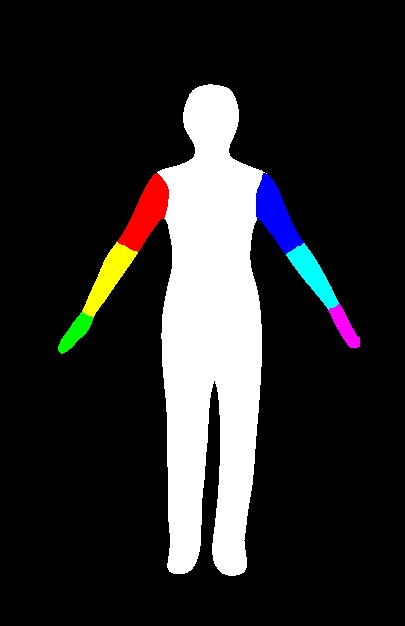}
    }
    \caption{Visual illustration of human body semantic maps: (a) The input image; (b) The DensePose map; (c) The human body semantic map of~\cite{wu2025low}; (d) Our proposed human body semantic map.}
    \label{fig:loose_semantic}
\end{figure}

\begin{figure}[htb]
  \centering
  
  \includegraphics[width=.9\linewidth]{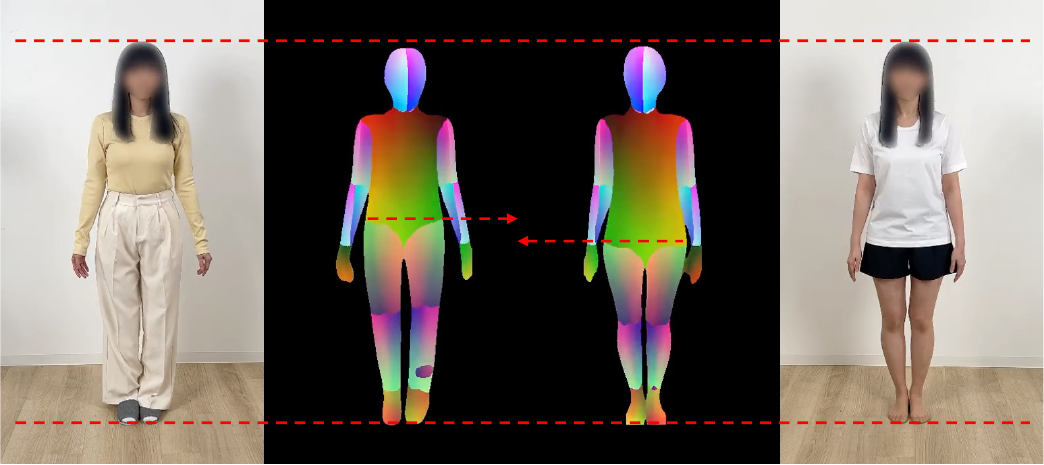}
  
  \caption{The boundary between the lower and upper body of the semantic map generated by DensePose~\cite{Guler2018DensePose} is sensitive to garment variation, even for tight-fitting garments.}
  \label{fig:loose_dp_sensitive}
\end{figure}

Unlike \cite{wu2025low}, which employs a simplified DensePose map as the human body semantic representation, our approach utilizes an even more simplified version, in which continuous UV coordinates are removed, as illustrated in~\cref{fig:loose_semantic}. This quantization enhances the learnability of the mapping from a garment-invariant representation to the human body semantic map. Following \cite{wu2025low}, we merged the lower and upper torso regions in the DensePose map. This was motivated by the observation that the boundary between these regions is not invariant to garment type, even for tight-fitting garments, as demonstrated in~\cref{fig:loose_dp_sensitive}.

We obtain the human body semantic map through a two-step process: First, we extract the garment-invariant representation from the input image. Next, we transform this garment-invariant representation into the human body semantic map using a per-person BodyMap network.

To generate the training data for the BodyMap network tailored to a specific person, we record a video of the person wearing a tight-fitting garment while performing predefined movements, as described in~\cite{wu2025low}. This person will later serve as the human model for per-garment dataset collection.

As illustrated in~\cref{fig:loose_overview}, we use the recorded video to generate paired garment-invariant representations and human body semantic maps to train the BodyMap network, whose architecture follows pix2pixHD~\cite{wang2018high}. The BodyMap network is then used to estimate the human body semantic map of the same person wearing a loose-fitting garment.

The BodyMap network is trained on a per-person basis and does not generalize across individuals. While this necessitates collecting a dedicated dataset for each human model, the process remains accessible, as it does not rely on specialized sensors to capture the body’s semantic map beneath loose-fitting garments. Moreover, since we typically use the same human model to collect multiple per-garment datasets, the trained BodyMap network can be efficiently reused.

\subsection{Dataset Generation for Loose-Fitting Garments}
For each garment item, we collect a garment-specific dataset, hereafter referred to as a per-garment dataset, to enable the training of garment-specific networks. Similar to the procedure in~\cite{wu2025low}, we record a video of the human model wearing the loose-fitting garment performing predefined movements. In addition, we record a separate video of the same human model wearing a tight-fitting garment and performing the same predefined movements, which is used to train the BodyMap network.

\begin{figure}[htb]
  \centering
  \includegraphics[width=.9\linewidth]{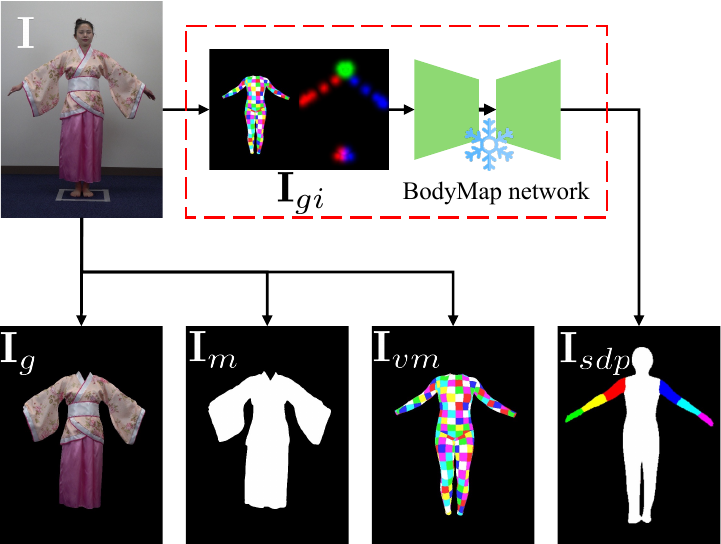}
  \caption{Per-garment dataset generation from recorded video frames of the human model. Unlike~\cite{wu2025low} that directly obtains the human body semantic map using DensePose~\cite{Guler2018DensePose}, we extract a garment-invariant representation and then transform it into the semantic map by our trained BodyMap network (red dashed box).}
  \label{fig:loose_dataset_gen}
\end{figure}

As illustrated in~\cref{fig:loose_dataset_gen}, given a frame $\mathbf{I}$ of the loose-fitting garment videos, we extract the garment image $\mathbf{I}_{g}$ and garment mask $\mathbf{I}_{m}$ using SAM2~\cite{ravi2024sam}. In addition, we generate the virtual measurement garment image $\mathbf{I}_{vm}$, which is also used to construct the garment-invariant representation $\mathbf{I}_{gi}$. By inputting the garment-invariant representation into the trained BodyMap network, we obtain the human body semantic map $\mathbf{I}_{sdp}$ robustly under loose-fitting garments.

After processing each frame of the loose-fitting garment video to obtain its corresponding $\mathbf{I}_{g}$, $\mathbf{I}_{m}$, $\mathbf{I}_{vm}$, and $\mathbf{I}_{sdp}$, we finish generating the per-garment dataset for the loose-fitting garment. We preserve the sequential order of each frame in the dataset to utilize temporal information for training.

\subsection{Training Pipeline of Virtual Try-On}
We follow~\cite{wu2025low} to utilize a hybrid person representation as the intermediate representation for garment synthesis. The hybrid person representation $\mathbf{I}_{hybrid}\in \mathbb{R}^{6 \times H \times W}$ is a combination of virtual measurement garment $\mathbf{I}_{vm}$ and human body semantic map $\mathbf{I}_{sdp}$, and can be formally represented as:

\begin{align} 
\mathbf{I}_{hybrid} = \mathbf{I}_{vm} \oplus \mathbf{I}_{sdp}
\end{align}
where $\mathbf{I}_{vm}$ captures the 3D pose of the human body which is crucial for $360^{\circ}$ virtual try-on, $\mathbf{I}_{sdp}$ provides pixel-level alignment guidance in 2D image space.

Unlike~\cite{wu2025low}, which trains the garment synthesis network on a per-frame basis, we train a Recurrent Garment Synthesis (ReGarSyn) network with sequential data to enhance the temporal consistency of the try-on results. We insert a lightweight ConvLSTM~\cite{shi2015convolutional} module into pix2pixHD~\cite{wang2018high} to obtain our ReGarSyn network.

\begin{figure}[htb]
  \centering
  \includegraphics[width=\linewidth]{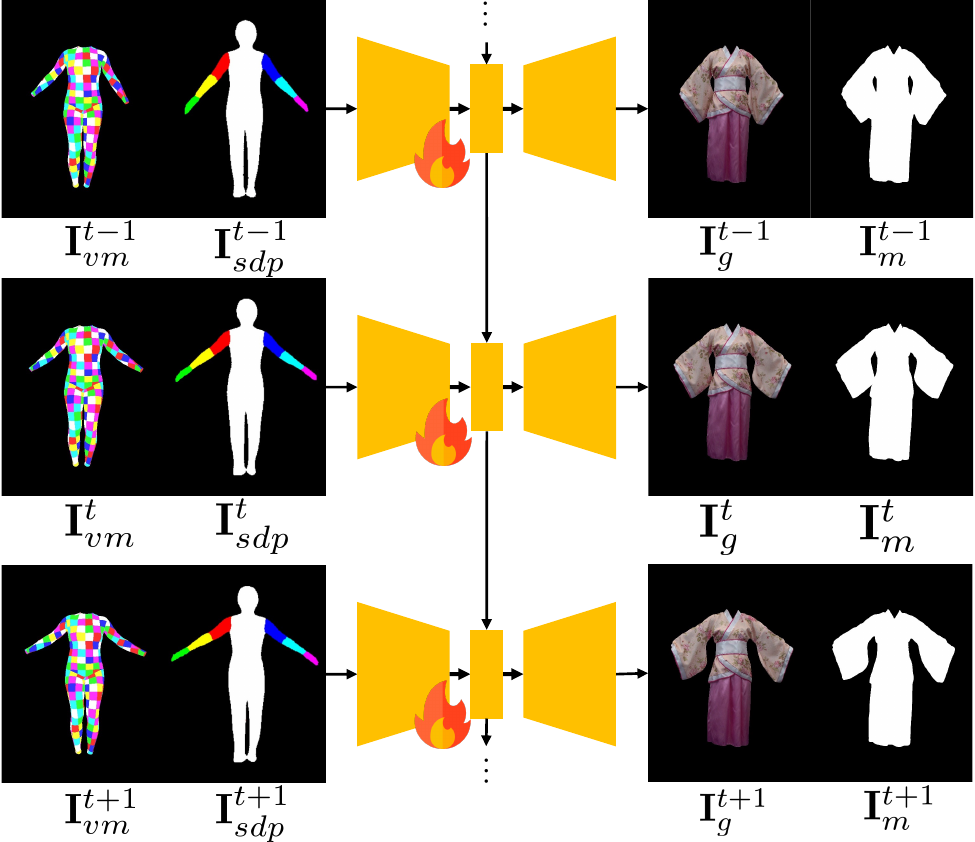}
  \caption{Detailed illustration of the ReGarSyn network training process.}
  \label{fig:loose_training}
\end{figure}

As illustrated in~\cref{fig:loose_training}, given an input sequence $\{ \mathbf{I}_{hybrid}^{1},\allowbreak \mathbf{I}_{hybrid}^{2},\allowbreak \ldots,\allowbreak \mathbf{I}_{hybrid}^{N} \}$ and a corresponding ground truth output sequence $\{ (\mathbf{I}_{g}^{1},\allowbreak\mathbf{I}_{m}^{1}),\allowbreak (\mathbf{I}_{g}^{2},\allowbreak\mathbf{I}_{m}^{2}),\allowbreak \ldots,\allowbreak (\mathbf{I}_{g}^{N},\mathbf{I}_{m}^{N}) \}$, the input and output of the ReGarSyn network at time step $t$ can be represented as:

\begin{align}
	\tilde{\mathbf{I}}_{g}^{t},\tilde{\mathbf{I}}_{m}^{t},\mathcal{C}^{t},\mathcal{H}^{t}=ReGarSyn(\mathbf{I}_{hybrid}^{t},\mathcal{C}^{t-1},\mathcal{H}^{t-1})
	\label{eq:update}
\end{align}
Here, $\tilde{\mathbf{I}}_{g}^{t}$ and $\tilde{\mathbf{I}}_{m}^{t}$ denote the synthesized garment image and mask at time step $t$, respectively. $\mathcal{C}$ and $\mathcal{H}$ represent the cell state and hidden state of the ConvLSTM module, capturing temporal dependencies across the sequence.

The loss function $\mathcal{L}$ can be defined as:

\begin{align}
	\mathcal{L}=\sum_{t=1}^{N} PerFrame(\tilde{\mathbf{I}}_{g}^{t} \oplus \tilde{\mathbf{I}}_{m}^{t},\mathbf{I}_{g}^{t} \oplus \mathbf{I}_{m}^{t})
\end{align}
where $PerFrame(\cdot,\cdot)$ denotes the per-frame loss function from pix2pixHD~\cite{wang2018high}. Explicit temporal consistency loss functions are not employed, as the supervision provided by temporally consistent ground truth data inherently ensures temporal coherence.

\subsection{Inference Pipeline of Virtual Try-On}

During inference, we assume the user is wearing a tight-fitting garment, which allows us to efficiently extract a semantic map of the human body using DensePose~\cite{Guler2018DensePose}. This assumption is practical and user-friendly, as it avoids the need for specialized wearable devices required by a prior approach~\cite{chong2021per}.

\begin{figure*}[htb]
\centering
\includegraphics[width=0.9\textwidth]{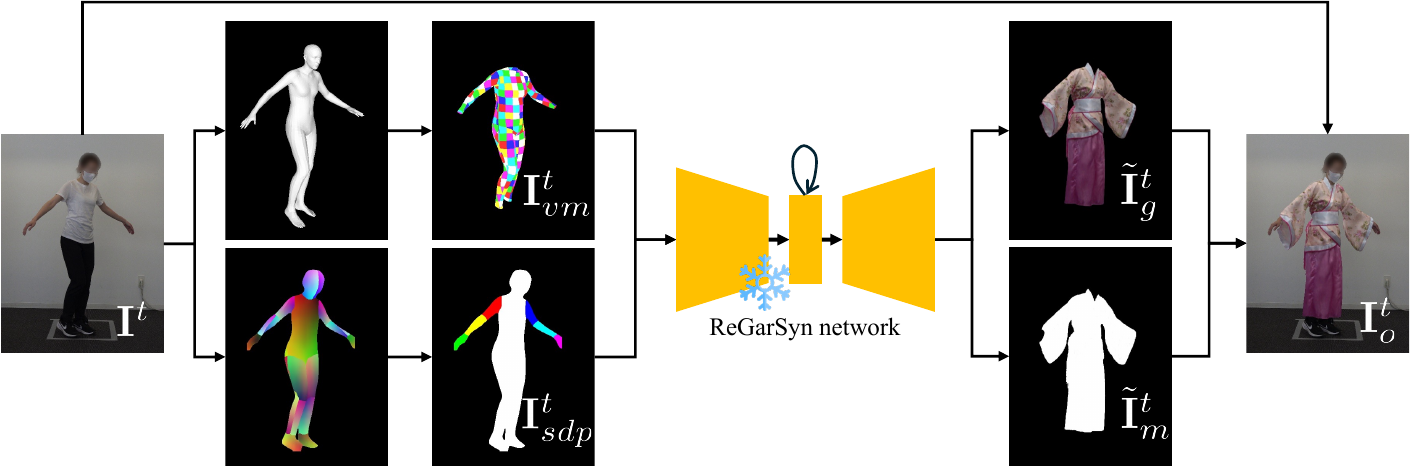}
\captionof{figure}{Detailed illustration of the inference pipeline for the ReGarSyn network.}
\label{fig:loose_inference}
\end{figure*}

As illustrated in~\cref{fig:loose_inference}, given an input frame $\mathbf{I}^{t}$ at time step $t$, we extract the virtual measurement garment $\mathbf{I}_{vm}^{t}$ using BEV~\cite{sun2022putting}, and obtain the human body semantic map $\mathbf{I}_{sdp}^{t}$ via DensePose~\cite{Guler2018DensePose}. Both processes are computationally efficient and suitable for real-time applications. Subsequently, as described in~\cref{eq:update}, we input them together with $\mathcal{C}^{t-1}$ and $\mathcal{H}^{t-1}$ from the previous time step to the trained ReGarSyn network to obtain the synthesized garment image $\tilde{\mathbf{I}}_{g}^{t}$ and mask $\tilde{\mathbf{I}}_{m}^{t}$. Finally, we composite them onto the input frame to obtain the try-on result $\mathbf{I}_o^{t}$:

\begin{align}
	\mathbf{I}_o^{t}=\mathbf{I}^{t}\odot (\mathbf{1}-\tilde{\mathbf{I}}_m^{t})+\tilde{\mathbf{I}}_g^{t}\odot \tilde{\mathbf{I}}_m^{t}
\end{align}
where $\odot$ denotes element-wise multiplication.

\section{Experiments}

\subsection{Experiment Setting}
\noindent \textbf{Per-garment datastes.} We collected per-garment datasets for four loose-fitting garments, including a hanfu, a down jacket, a hanbok, and a dress, using the same human model. Each dataset contains approximately 3,000 images. Additionally, we recorded a video of the same human model wearing a tight-fitting garment and performing similar movements for evaluation purposes.

\noindent \textbf{Implementation details.} We set the resolution of the BodyMap network to $512 \times 384$ and the ReGarSyn networks to $576 \times 432$. For each garment, we trained a per-garment model using the corresponding per-garment dataset. We trained each model on an NVIDIA RTX 4090 GPU for $40$ epochs, with a learning rate of $2\times10^{-4}$. During training, we randomly sampled clips ranging in length from 8 to 60 frames. We use the Adam optimizer with $\beta_1=0.5$ and $\beta_2=0.999$. The training duration for a garment-specific network was approximately $12$ hours.

\noindent \textbf{Comparison setting.} We conduct both qualitative and quantitative evaluations of our method, using the prior per-garment virtual try-on approach~\cite{wu2025low} as the baseline for comparison. Additionally, we compare our results with OOTDiffusion~\cite{xu2024ootdiffusion} and ViViD~\cite{fang2024vivid}, which represent state-of-the-art image-based and video-based virtual try-on methods, respectively. Due to mismatches in input and output formats, retraining OOTDiffusion and ViViD on our datasets is not feasible, limiting the fairness of direct comparisons. Nonetheless, we include these methods to demonstrate that their reliance on a universal network for generalization across all garment types proves inadequate in practice, especially for rarely seen garments.

\subsection{Qualitative Evaluation}
\label{sec:loose_qualitative}
\begin{figure}[htb]
  \centering
  \includegraphics[width=\linewidth]{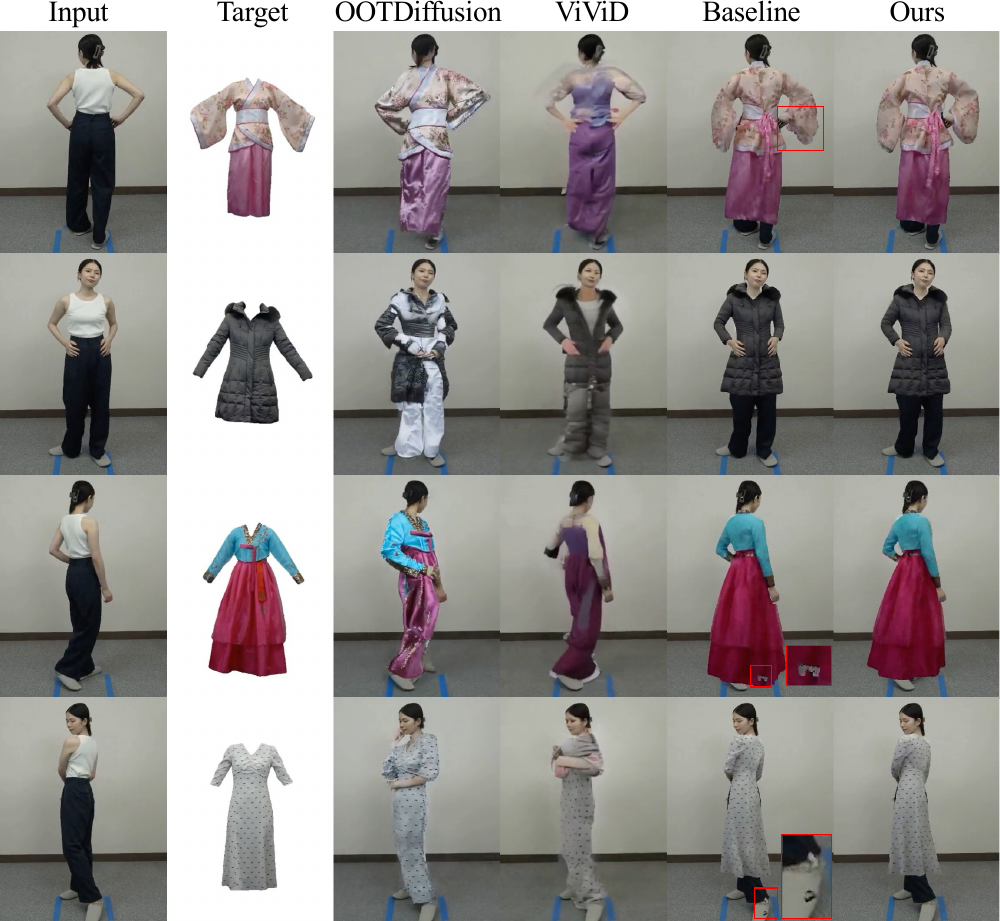}
  \caption{Qualitative comparison of our method against OOTDiffusion~\cite{xu2024ootdiffusion}, ViViD~\cite{fang2024vivid}, and the baseline method~\cite{wu2025low}.}
  \label{fig:loose_qualitative}
\end{figure}

We present a qualitative comparison of image quality among OOTDiffusion~\cite{xu2024ootdiffusion}, ViViD~\cite{fang2024vivid}, the baseline method~\cite{wu2025low}, and our method in~\cref{fig:loose_qualitative}. Both OOTDiffusion and ViViD produce poor results for these loose-fitting garments, which supports our assertion that training a universal network on frequently seen garments cannot generalize well to all types of garments.

In the results produced by the baseline method, noticeable flaws appear in regions with loose-fitting garments—such as the wide sleeves of the hanfu (first row), the skirt hems of the hanbok (third row), and the hems of the dress (fourth row). These artifacts support our assertion that the baseline method struggles with discrepancies in input data distribution between the training and inference stages due to occlusions introduced by loose-fitting garments. While the down jacket (second row) does not exhibit such artifacts due to its relatively tight-fitting structure, it reveals temporal inconsistencies attributable to the baseline’s frame-by-frame garment synthesis strategy, as further evidenced by the quantitative results in~\cref{sec:loose_quantitative}. For a more intuitive video comparison, please refer to the supplementary video.

\begin{figure}[htb]
  \centering
  \includegraphics[width=\linewidth]{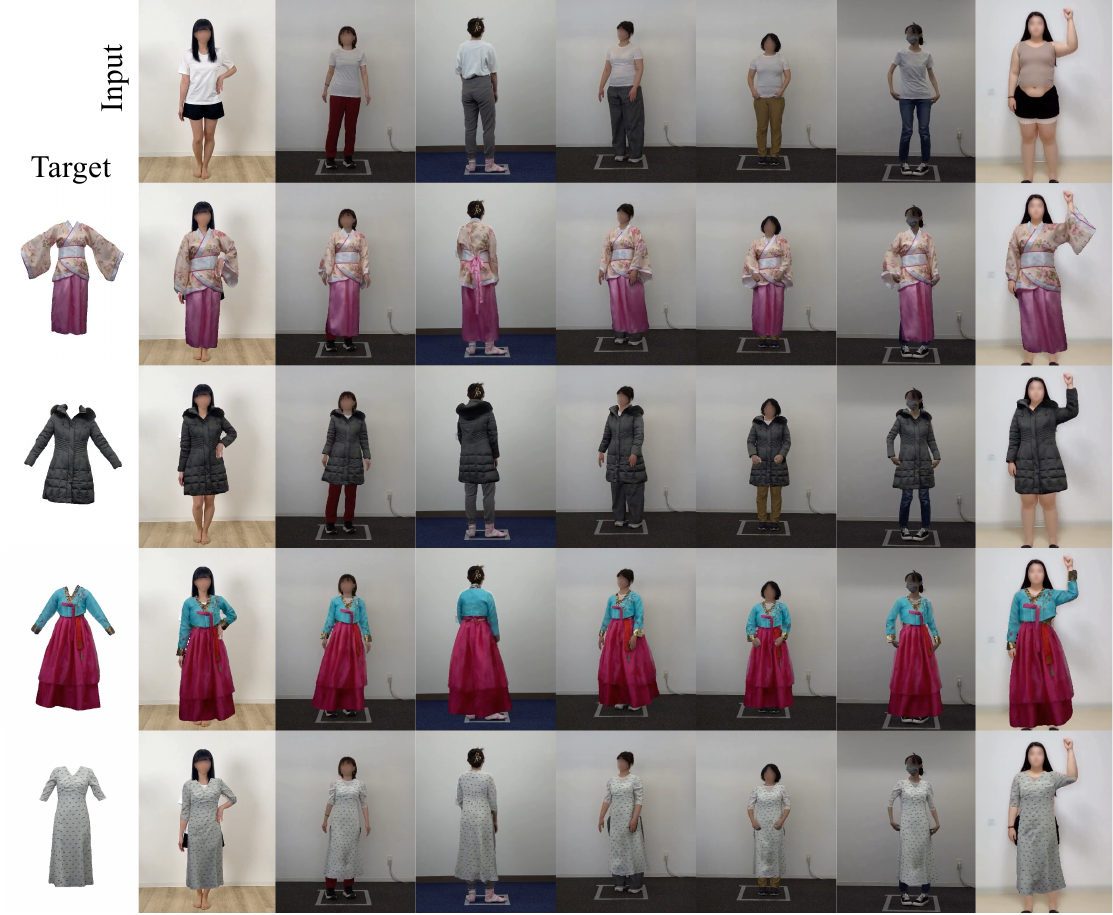}
  \caption{Virtual try-on results for unseen body shapes. Our ReGarSyn networks trained on per-garment datasets collected by a single human model demonstrate strong generalization to unseen body shapes.}
  \label{fig:loose_body_generalize}
\end{figure}

Furthermore, to demonstrate the generalization ability of our ReGarSyn networks trained on per-garment datasets collected by a single human model, we present the virtual try-on results for unseen body shapes in~\cref{fig:loose_body_generalize}. This figure indicates that our strategy of using only one human model for each per-garment dataset is sufficient, as the ReGarSyn network successfully generalizes to a wide range of body shapes.

\subsection{Quantitative Evaluation}
\label{sec:loose_quantitative}

Since collecting paired ground truth data for loose-fitting garment virtual try-on is impractical, we adopt evaluation metrics that do not rely on such data. Specifically, we use Kernel Inception Distance (KID) and Fréchet Inception Distance (FID) to assess image quality. To evaluate both image quality and temporal consistency, we employ Video Fréchet Inception Distance (VFID), which leverages the I3D backbone~\cite{carreira2017quo}. These metrics are computed using two videos: one showcasing the generated virtual try-on result and the other depicting the same person wearing the actual loose-fitting garment. Although the videos exhibit similar human motion, they are not strictly synchronized.

We report the quantitative results for four loose-fitting garments individually in~\cref{tab:loose_quantitative}. As shown in the table, both OOTDiffusion~\cite{xu2024ootdiffusion} and ViViD~\cite{fang2024vivid} perform significantly worse than the baseline and our proposed method, which is consistent with the qualitative observations discussed in~\cref{sec:loose_qualitative}. Although the baseline occasionally surpasses our method in certain image quality metrics, our approach achieves comparable scores overall and consistently outperforms all others in VFID across all garments. This highlights the strength of our method in maintaining both temporal consistency and visual fidelity.

\begin{table*}[htbp]
    \centering
    \footnotesize 
    \caption{Quantitative comparison with other approaches and ablation study on the ConvLSTM module.}
    \label{tab:loose_quantitative}
    \begin{tabular}{@{}l|ccc|ccc|ccc|cccccc@{}}
    \toprule
    \multirow{2}{*}{Method} & \multicolumn{3}{c}{Hanfu} & \multicolumn{3}{c}{Down jacket} & \multicolumn{3}{c}{Hanbok} & \multicolumn{3}{c}{Dress} \\
    \cmidrule(lr){2-4} \cmidrule(lr){5-7} \cmidrule(lr){8-10} \cmidrule(lr){11-13}
    & KID$\downarrow$ & FID$\downarrow$ & VFID$\downarrow$  & KID$\downarrow$ & FID$\downarrow$ & VFID$\downarrow$  & KID$\downarrow$ & FID$\downarrow$ & VFID$\downarrow$  & KID$\downarrow$ & FID$\downarrow$ & VFID$\downarrow$  \\
    \midrule
    OOTDiffusion~\cite{xu2024ootdiffusion} & 0.073 & 77.27 & 1.138 & 0.180 & 154.43 & 1.234 & 0.257 & 187.30 & 1.359 & 0.116 & 100.47 & 1.036 \\
    ViViD~\cite{fang2024vivid} & 0.190 & 153.88 & 1.266 & 0.200 & 173.29 & 1.115 & 0.439 & 285.40 & 1.254 & 0.125 & 105.57 & 0.996 \\
    Baseline~\cite{wu2025low} & 0.043 & 51.25 & 1.041 & 0.102 & 97.81 & 1.027 & 0.022 & 28.85 & 0.942 & \textbf{0.071} & \textbf{66.75} & 0.992 \\

    Ours w/o ConvLSTM & 0.037 & 44.24 & 1.027 & 0.102 & 98.41 & 1.029 & \textbf{0.020} & \textbf{26.92} & 0.933 & 0.072 & 67.00 & 0.981 \\
    
    Ours & \textbf{0.034} & \textbf{42.88} & \textbf{1.025} & \textbf{0.099} & \textbf{96.86} & \textbf{1.024} & 0.021 & 27.01 & \textbf{0.930} & 0.073 & 67.26 & \textbf{0.980} \\
    \bottomrule
    \end{tabular}
    
\end{table*}

\subsection{Frame Rate Comparison}
\label{sec:loose_framerate}

We compare the frame rate of our method against OOTDiffusion~\cite{xu2024ootdiffusion}, ViViD~\cite{fang2024vivid}, the baseline approach~\cite{wu2025low}, and a variant of our model that shares the same architecture but excludes the ConvLSTM module and is trained on a per-frame basis ("ours w/o ConvLSTM"). This variant is included to evaluate the impact of the ConvLSTM module on frame rate performance. All frame rates are measured using an NVIDIA RTX 4090 GPU.

As shown in~\cref{tab:loose_efficiency}, both OOTDiffusion and ViViD exhibit frame rates that fall short of real-time requirements. The baseline achieves the highest frame rate, while the "ours w/o ConvLSTM" variant, which incorporates additional processing for the DensePose map, performs slightly slower. Our proposed method integrates a ConvLSTM module, resulting in a $14\%$ reduction in frame rate. Despite this, it maintains a frame rate of 10.5 frames per second, which remains adequate for real-time use.

\setlength{\tabcolsep}{4pt}
\begin{table}[hbt!]
\begin{center}
\caption{Comparison of frame rates. The introduction of the ConvLSTM module results in a slight decrease in frame rate but remains sufficient for real-time performance use.}
\label{tab:loose_efficiency}
\begin{tabular}{lc}
\toprule[1pt]
Method  & Frame rate (fps) $\uparrow$  \\
\noalign{\smallskip}
\hline
\noalign{\smallskip}
OOTDiffusion~\cite{xu2024ootdiffusion}  & 0.32  \\
ViViD~\cite{fang2024vivid} & 1.32 \\
Baseline~\cite{wu2025low} & \textbf{12.38}  \\
Ours w/o ConvLSTM & 12.15  \\
Ours & 10.50  \\
\toprule[1pt]
\end{tabular}
\end{center}
\end{table}
\setlength{\tabcolsep}{1.4pt}

\subsection{Ablation Study on Garment-Invariant Representation}

\begin{figure}[htb]
    \centering
    \subfloat[HM]{%
        \includegraphics[width=0.23\linewidth]{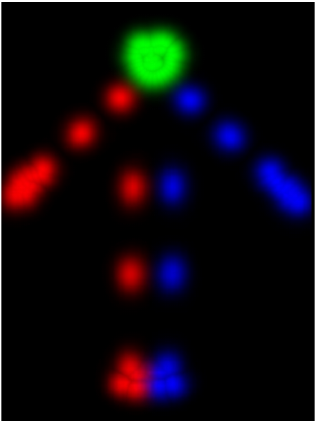}
    }
    \hspace{0cm}
    \subfloat[SHM]{%
        \includegraphics[width=0.23\linewidth]{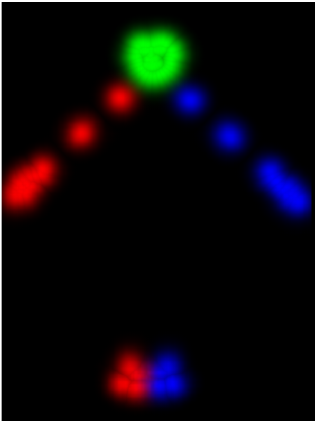}
    }
    \hspace{0cm}
    \subfloat[VM]{%
        \includegraphics[width=0.23\linewidth]{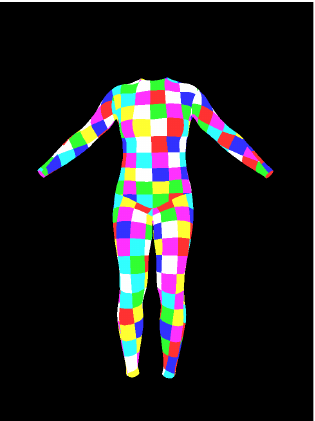}
    }
   
    \caption{Three alternative intermediate representations used in ablation study for human body semantic map estimation.}
    \label{fig:loose_gi_ablation}
\end{figure}

To demonstrate the necessity of our proposed garment-invariant representation and its components, we conduct ablation studies with the following alternative approaches for human body semantic map estimation:

\begin{itemize}
	\item \textbf{DP}: Directly using DensePose~\cite{Guler2018DensePose} to obtain the human body semantic map without any intermediate representation. This is to demonstrate the benefit of our proposed garment-invariant representation.
	\item \textbf{HM}: Using heatmaps without removing any joints as the intermediate representation (\cref{fig:loose_gi_ablation}(a)). This is to show the significance of removing garment-sensitive joints.
	\item \textbf{SHM}: Using heatmaps without garment-sensitive joints as the intermediate representation (\cref{fig:loose_gi_ablation}(b)). This reveals the importance of the virtual measurement garment~\cite{wu2024virtual}.
	\item \textbf{VM}: Using a virtual measurement garment~\cite{wu2024virtual} as the intermediate representation (\cref{fig:loose_gi_ablation}(c)). This demonstrates the importance of joint heatmaps.
\end{itemize}

We lack ground truth data for estimating human body semantic maps under loose-fitting garments, as obtaining paired images of the same person in identical poses wearing both tight-fitting and loose-fitting clothing is not feasible. Therefore, we can only conduct a qualitative comparison. The metrics used in~\cref{sec:loose_quantitative} are unsuitable here since they cannot accurately reflect how well the semantic maps align with the human body.

\begin{figure}[htb]
  \centering
  \includegraphics[width=.9\linewidth]{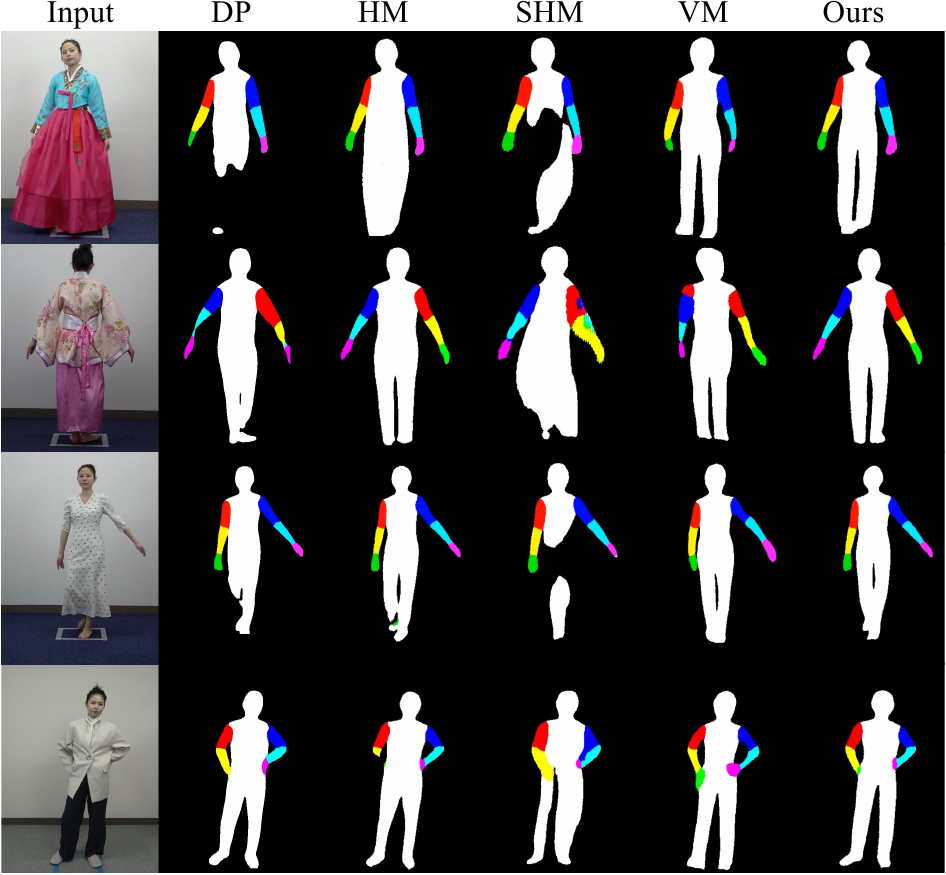}
  \caption{Ablation study on garment-invariant representation. \textbf{DP}: The semantic map estimated by DensePose~\cite{Guler2018DensePose} is not robust against loose-fitting garments. \textbf{HM}: Using joint heatmaps that include garment-sensitive joints leads to the inflation of the semantic map. \textbf{SHM}: Using joint heatmaps without garment-sensitive joints fails to provide adequate information for semantic map estimation. \textbf{VM}: Using solely virtual measurement garments fails to provide pixel-level accurate guidance for semantic maps.}
  \label{fig:loose_ablation}
\end{figure}

As shown in~\cref{fig:loose_ablation}, our proposed garment-invariant representation enables reliable human body semantic map estimation. The semantic maps estimated by DensePose suffer from occlusion, e.g., the lower body region is missing due to the occlusion of the hanbok skirt (first row, DP column). Using joint heatmaps that include garment-sensitive joints results in inflated semantic maps (HM column). Utilizing joint heatmaps without garment-sensitive joints fails to provide sufficient information for semantic map estimation, resulting in distorted and incomplete results (SHM column). Relying solely on the virtual measurement garments does not provide pixel-level guidance for semantic map estimation, resulting in limbs being placed in incorrect regions (second row, VM column).

\subsection{Ablation Study on ConvLSTM Module}

We evaluate the effectiveness of the ConvLSTM module in enhancing temporal consistency through both qualitative and quantitative comparisons with a variant without ConvLSTM ("ours w/o ConvLSTM"), as described in~\cref{sec:loose_framerate}.

\begin{figure}[htb]
  \centering
  \includegraphics[width=\linewidth]{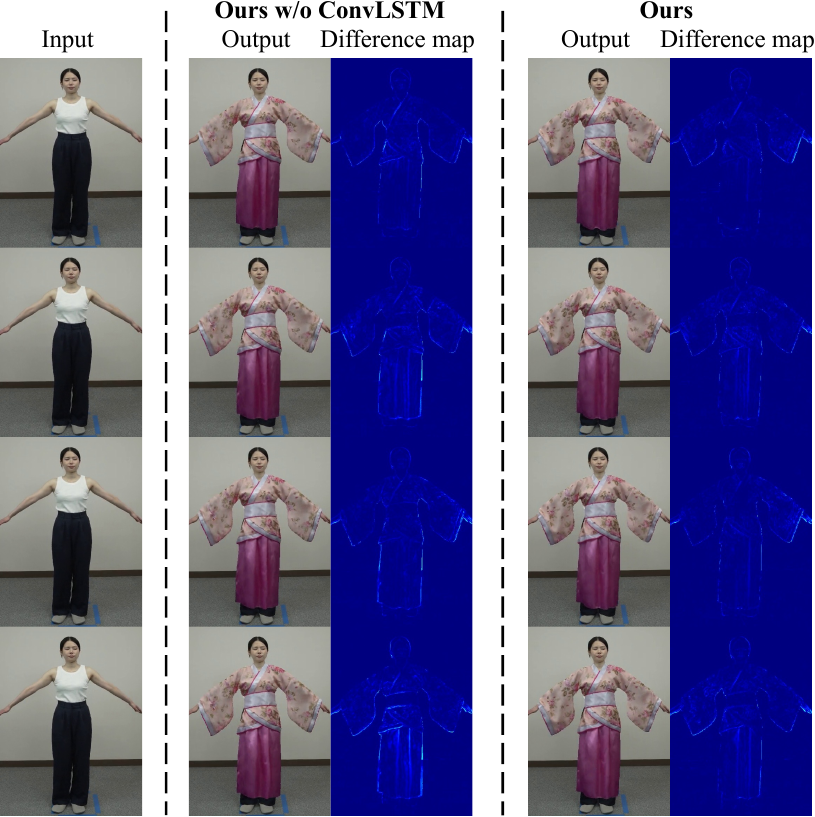}
  \caption{Ablation study on ConvLSTM module. The difference map visualizing absolute differences between consecutive output frames shows that the ConvLSTM module enhances temporal consistency effectively.}
  \label{fig:loose_temporal_ablation}
\end{figure}

To better illustrate the presence of jittering artifacts, we visualize the absolute differences between consecutive output frames using heatmaps, as shown in~\cref{fig:loose_temporal_ablation}. These visualizations reveal that the "ours w/o ConvLSTM" variant exhibits noticeable jittering artifacts, particularly along the boundary of the loose skirt, which tends to vary in appearance even when the human remains relatively static. In contrast, our method, which incorporates the ConvLSTM module, achieves notably improved temporal consistency across frames. For a more intuitive comparison, please refer to the supplementary video.

\section{Limitations and Future Work}

In addition to the common limitations mentioned by~\cite{wu2025low}, such as the original garment not being removed, and inaccurate composition masks, we would like to highlight the following limitations and their future work:

\begin{description}[style=unboxed,leftmargin=0cm]
	\item[The necessity of per-garment dataset and training.] Our method requires collecting a dedicated dataset and training a separate network for each garment, which poses scalability challenges. To address this issue, future research could explore a per-garment-type strategy, where a single dataset is collected and a single network is trained for each category of similar garments rather than for each specific item. This approach would substantially reduce training time and improve scalability. 
	\item[Assumption of tight-fitting garments during inference.] We assume that users are wearing tight-fitting garments when interacting with our virtual try-on system. The current implementation does not accommodate scenarios where users wear loose-fitting garments, which may lead to performance degradation due to occlusion effects. To overcome this limitation, future research could investigate real-time human body representations that are robust to the presence of loose-fitting garments.
    \item[Inability to convey actual size and fit.] While our method generates visually plausible results for a wide range of body shapes and provides useful styling references (\cref{fig:loose_body_generalize}), it does not validate the suitability of garment sizing due to the lack of actual user body measurements. Although some general image-based methods~\cite{chen2023size,chen2025sico} incorporate garment size information, they fall short of accurately reflecting the real physical properties of a specific garment. A promising direction for future research is developing a garment synthesis framework that incorporates user-provided body measurements and utilizes per-garment datasets annotated with body measurement metadata. This approach would enable more accurate and personalized virtual try-on experiences.
    \item[Limitation in modeling transient physical dynamics.] Although our recurrent framework incorporates temporal information by conditioning garment synthesis on preceding frames, its primary objective is to enforce inter-frame consistency rather than to model the transient physical dynamics of loose-fitting garments, such as fluttering. The datasets we collected using the method of~\cite{wu2025low} comprise sequences of humans rotating in place, offering inadequate motion diversity for capturing the complex physical behaviors of loose garments. To address this limitation, a novel dataset collection method that introduces a broader range of motion while remaining accessible needs to be explored.
    
\end{description}
    
\section{Conclusion}
We introduce a novel per-garment approach that is the first to enable real-time and temporally coherent virtual try-on for loose-fitting garments. To enhance the quality of synthesized images, we propose a garment-invariant representation that facilitates accurate estimation of human body semantic maps under loose-fitting garments during per-garment dataset generation. This representation effectively mitigates the distribution gap between the training and inference stages of our garment synthesis network, resulting in significantly improved visual fidelity. To ensure temporal consistency, we propose a recurrent framework that leverages temporal dependencies for garment synthesis. Our inference pipeline demonstrates sufficient efficiency to support real-time virtual try-on and can process input frames of arbitrary length. Through extensive qualitative and quantitative evaluations, we demonstrate that our method outperforms existing approaches in both image quality and temporal consistency. We also conduct ablation studies to validate the effectiveness of the garment-invariant representation and the recurrent synthesis framework.


\bibliographystyle{ACM-Reference-Format}
\bibliography{paper.bib}
\clearpage




\end{document}